\documentclass[12pt]{article}
\usepackage[xdvi]{graphicx}
\usepackage{amssymb}
\makeatletter
\@addtoreset{equation}{section}
\makeatother

\newcommand{\bea}{\begin{eqnarray}}
\newcommand{\eea}{\end{eqnarray}}

\begin{document}
\begin{titlepage}
\begin{flushright}
HIP-2008-20/TH \\
\end{flushright}

\vspace*{5cm}

\begin{center}{\Large\bf
The lightest Higgs boson mass in effective field theory
with bulk and brane supersymmetry breaking
}
\end{center}

\vspace{1ex}

\begin{center}
{\large 
Nobuhiro Uekusa}
\end{center}
\begin{center}
{\it Department of Physics, 
University of Helsinki  \\
and Helsinki Institute of Physics, \\
P.O. Box 64, FIN-00014 Helsinki, Finland} \\
\textit{E-mail}: nobuhiro.uekusa@helsinki.fi
\end{center}


\vspace{1ex}

\begin{abstract}
We study the mass of the lightest Higgs boson
in an effective Kaluza-Klein theory with
two sources of supersymmetry breaking.
When stop mass matrices are non-diagonal 
with respect to Kaluza-Klein modes and each element is not small
in unit of the inverse compactification radius,
a very small eigenvalue can be accommodated 
in the diagonalized basis.
We show that in the model with the two sources,
the Higgs boson mass can receive suitable corrections
for various compactification radii.

\end{abstract}
\end{titlepage}



\newpage
\pagenumbering{arabic}

\section{Introduction}

There has been much attention to the values of the masses of Higgs bosons.
The Higgs field of the Standard Model has a classical potential composed of
quadratic and quartic self-couplings.
The mass receives radiative corrections from loop diagrams.
If the Higgs field couples to a fermion with a 
dimensionless Yukawa coupling constant
and an ultraviolet moment cutoff is employed to regulate the loop integral,
the correction to the mass squared yields a quadratic term of the cutoff.
There are similar contributions from the virtual effects of heavy fields,
as emphasized in the introduction~\cite{Martin:1997ns}.
When there is a heavy complex scalar field
$\Phi$ coupled to the Higgs field $H$ with the coupling
$-\lambda_\Phi|H|^2|\Phi|^2$, the Higgs field
receives a correction to the squared mass dependent on the cutoff,
\bea
   \Delta m_H^2 =
    {\lambda_\Phi\over 16\pi^2}\left[
    \Lambda^2
    -2m_\Phi^2\textrm{ln}\left({\Lambda\over m_\Phi}\right)
    \right] .
    \label{mhboson}
\eea
In addition to a quadratic term of the ultraviolet momentum cutoff $\Lambda$,
there is a quadratic term of the heavy field mass $m_\Phi$.
Heavy fermions also contribute similar corrections.
In the light of the sensitivity of the Higgs boson mass to the masses of 
heavy fields, 
the effects on the Standard Model
are not decoupled to heavy fields.

Once supersymmetry is assumed,
such terms sensitive to the ultraviolet momentum cutoff and heavy field masses
are canceled between bosonic and fermionic contributions.
Because superparticles have not been detected, supersymmetry must be broken.
In order to keep cancellation of the quadratic term on the cutoff,
supersymmetry breaking needs to be soft.
After supersymmetry is broken,
the divergent correction to the Higgs boson mass squared is of a logarithmic form
\bea
  \Delta m_H^2 =m_{\textrm{\scriptsize sp}}^2
    \left[{\lambda\over 16\pi^2}
      \ln \left({\Lambda\over m_{\textrm{\scriptsize sp}}}\right)
      \right] .
 \label{mhsoft}
\eea
Here $m_{\textrm{\scriptsize sp}}$ is the largest mass scale associated with
the mass splittings between bosons and fermions
and $\lambda$ stands for dimensionless couplings schematically.
In the Minimal Supersymmetric Standard Model (MSSM)
with $\Lambda$ of the Planck scale and $\lambda$ of order
${\cal O}(1)$,
the $m_{\textrm{\scriptsize sp}}$ is at the most 
1~TeV so that the correction $\Delta m_H^2$ is not too large compared to the 
electroweak breaking scale.

To achieve mass splittings is of great interest
in its own right.
Supersymmetry must be broken in a way with no
supersymmetric flavor problem.
One candidate is to spatially separate 
the MSSM fields from the source of 
supersymmetry breaking~\cite{Mirabelli:1997aj}%
-\cite{Chacko:1999mi}, as the mediation in bulk spacetime does not distinguish flavor.
When the radius of an extra dimension is denoted as $R$,
the mass splitting between a bulk field and its supersymmetric partner
can be of the order of $R^{-1}$ with a factor of order ${\cal O}(1)$.
This type of mass splitting is obtained by 
$F$-term supersymmetry breaking on a brane~%
\cite{Mirabelli:1997aj}\cite{Marti:2001iw}%
-\cite{DiClemente:2002qa}
or independently by Scherk-Schwarz supersymmetry breaking~%
\cite{Scherk:1978ta}%
-\cite{Diego:2006py}.
If $R^{-1}\sim$ 1~TeV,
corrections to the Higgs mass would be significant.
In a Kaluza-Klein approach, 
bulk fields are decomposed in zero mode and Kaluza-Klein modes.
Kaluza-Klein modes contain many heavy fields and each provides 
corrections of the form given in Eq.~(\ref{mhboson}). 
In models with softly broken supersymmetry, corrections would have the 
form (\ref{mhsoft}). Here
each of Kaluza-Klein modes as well as zero mode
acquires the mass splitting between bosonic and fermionic fields. 
Because the contribution from each mode is summed, 
supersymmetry breaking with extra dimensions 
may increase the corrections to Higgs boson masses.
In the context of extra dimensions, effective potentials and
Higgs boson masses have been widely studied~%
\cite{DiClemente:2002qa}\cite{Hatanaka:1998yp}%
-\cite{Lim:2007jv}.

Both of the two ways of supersymmetry breaking above 
can provide the mass splitting of the same order.
On the other hand,
there is a difference about the structure of Kaluza-Klein mode between them.
For the Scherk-Schwarz supersymmetry breaking,
supersymmetry-breaking 
mass matrices in the Lagrangian are diagonal with respect to Kaluza-Klein mode and
$F$-term supersymmetry breaking on a brane leads to non-diagonal 
supersymmetry-breaking mass matrices.
In the latter case each component of the mass matrices 
is of the order of $R^{-1}$
and in the diagonalized basis the eigenvalues are also of the order of
$R^{-1}$.

In the case with the Scherk-Schwarz twist, $N=1$ supersymmetry
can be locally preserved on each boundary. Thus supersymmetric action can be given at each boundary and it may be locally broken by some brane effects. Such a combined supersymmetry breaking is physically inequivalent to that of only one supersymmetry-breaking source because of distinct mass spectrum. However, its possibility has not been considered in the literature.
A point to be examined is whether
supersymmetry-breaking 
mass matrices assembled of the two possible sources about
of order ${\cal O}(1)$ can 
give rise to at least one small eigenvalue in the diagonalized basis.
If very small eigenvalues are generated, the effect of radiative corrections
might be too small contributions to explain the Higgs boson mass in an allowed region for $R^{-1}\sim 1$~TeV.
Instead
it may be possible that the resulting Higgs mass amounts 
to being corrected for a large $R^{-1}$ such as the 
grand unification scale
due to the mass splitting made of $R^{-1}$ multiplied by the small factor.

We study the corrections to the mass of the lightest Higgs boson
in a Kaluza-Klein effective field theory.
In a simple model, it is explicitly shown how the contribution from the mass splitting of each
Kaluza-Klein level is summed.
In a model with supersymmetry broken by Scherk-Schwarz twists and 
$F$-terms on a brane,
we find that small eigenvalues for mass matrices 
can be accommodated. 
We show that for the two supersymmetry-breaking sources,
quantum loop corrections with a variety of 
compactification scales can relax the upper bound to
the mass of the lightest Higgs boson.

The paper is organized as follows.
In Section~\ref{lesthiggs}, our notation for the lightest Higgs boson
is given.
In Section~\ref{rad}, a picture for obtaining
radiative corrections to the lightest Higgs boson
is shown for the four-dimensional case and Kaluza-Klein case.
In Section~\ref{SS}, our model with Scherk-Schwarz twists and boundary $F$-term
is presented. We calculate the diagonalization of 
mass matrices with respect to Kaluza-Klein modes.
A conclusion is made in Section~\ref{conclusion}.

\section{The Higgs bosons in a mass-eigenstate basis \label{lesthiggs}}

In this section, we summarize our notation for the lightest Higgs boson
following Ref.~\cite{Martin:1997ns, Weinberg}.
The Higgs scalar fields in the MSSM consist of two complex SU(2)${}_L$-doublets
$H_1$ and $H_2$ which in total have 8 real scalar degrees of freedom
\bea
 H_1&\!\!\!=\!\!\!&
\left(\begin{array}{c}
 H_1^0 \\
 H_1^- \\
 \end{array}\right)
 =\left(\begin{array}{c}
 H_d^0 \\
 H_d^- \\
 \end{array}\right) ,
 \qquad
 (1,2,-\textrm{${1\over 2}$}) ,
\\
  H_2 &\!\!\!=\!\!\!&
\left(\begin{array}{c}
 H_2^+ \\
 H_2^0 \\
 \end{array}\right)
 =\left(\begin{array}{c}
 H_u^+ \\
 H_u^0 \\
 \end{array}\right),
 \qquad
 (1,2,+\textrm{${1\over 2}$}) .
 \eea
Here the quantum numbers for SU(3)$\times$ SU(2)${}_L$ $\times$ U(1)${}_Y$
are indicated by the numbers in the parentheses.
The superpartners are the spin 1/2 higgsinos
\bea
 \left(\begin{array}{c}
 \tilde{H}_1^0 \\
 \tilde{H}_1^- \\
 \end{array}\right), ~
\left(\begin{array}{c}
 \tilde{H}_2^+ \\
 \tilde{H}_2^0 \\
 \end{array}\right) .
\eea
We express the superpartners of the Standard Model fields 
by putting a tilde on the corresponding letter.
The potential of the Higgs scalar fields is
\bea
  V&\!\!\!=\!\!\!& (|\mu|^2+m_1^2)|H_1|^2+(|\mu|^2+m_2^2)|H_2|^2
    +\left[b H_1 H_2 +\textrm{H.c.}\right]
\nonumber
\\
 &&+{1\over 8}g^2 (H_2^\dag \sigma^a H_2 +H_1^\dag \sigma^a H_1)^2
   +{1\over 8} g'{}^2(|H_2|^2-|H_1|^2)^2
\nonumber
\\
 &\!\!\!=\!\!\!& (|\mu|^2+m_1^2) (|H_1^0|^2+|H_1^-|^2)
 +(|\mu|^2+m_2^2)(|H_2^+|^2+|H_2^0|^2)
\nonumber
\\
 &&
+\left[b(H_1^- H_2^+ - H_1^0 H_2^0) +\textrm{H.c.}\right]
\nonumber
\\
 &&+{1\over 8}(g^2+g'{}^2)(|H_2^+|^2+|H_2^0|^2
 -|H_1^0|^2-|H_1^-|^2)^2
 +{1\over 2}g^2 |H_2^+ H_1^{0*}
  +H_2^0 H_1^{-*}|^2 ,
  \label{hpot}
\eea
where $g$ and $g'$ denote SU(2)${}_L$ and U(1)${}_Y$ gauge coupling
constants, respectively,
$\mu$ is a supersymmetric mass and $m_1$, $m_2$ and $b$
are supersymmetry-breaking coupling constants.

For $H_2^+=H_1^-=0$, the Higgs potential is
\bea
  V&\!\!\!=\!\!\!&(|\mu|^2 +m_1^2)|H_1^0|^2 
  +(|\mu|^2 +m_2^2)|H_2^0|^2
    -\left[bH_1^0 H_2^0 +\textrm{H.c.}\right]
\nonumber
\\
  &&+{1\over 8}(g^2+{g'}^2)(|H_2^0|^2 -|H_1^0|^2)^2 .
\eea
The equilibrium conditions
$\partial V/\partial H_2^0=\partial V/\partial H_1^0=0$ at
$H_1^0=v_1,H_2^0=v_2$ lead to
\bea
    |\mu|^2 +m_2^2 -b \cot \beta -(m_Z^2/2) \cos 2\beta =0 , 
    \label{mumz2}
\\
    |\mu|^2 +m_1^2 -b \tan \beta +(m_Z^2/2) \cos 2\beta =0 , 
  \label{mumz}
\eea
respectively. Here 
\bea
   \tan\beta ={v_2\over v_1} ,\quad
   m_W^2={1\over 2}g^2 (v_1^2+v_2^2) ,\quad
   m_Z^2={1\over 2}(g^2+{g'}^2)(v_1^2+v_2^2) .
\eea
Eqs.~(\ref{mumz2}) and (\ref{mumz}) 
require
\bea
  b^2 > (|\mu|^2+m_1^2)(|\mu|^2+m_2^2) .
\eea  

When the electroweak symmetry is broken,
three among the eight degrees of freedom 
of the Higgs scalar fields are the would-be Nambu-Goldstone bosons
$G^0$ and $G^{\pm}$
which become the longitudinal modes of the $Z^0$ and $W^{\pm}$
massive vector bosons.
The remaining five eigenstates are classified in
three electrically-neutral fields and two electrically-charged fields.
Electrically-neutral fields consist of two CP-even
scalars $h^0$ and $H^0$
and one CP-odd scalar $A^0$.
Electrically-charged fields are one scalar $H^+$ with charge $+1$
and its conjugate scalar $H^-$ with charge $-1$.
Here $h^0$ is lighter than $H^0$. The charged scalars are
subject to $G^-=G^{+*}$, $H^-=H^{+*}$.
The gauge-eigenstate fields can be expressed in terms of the mass-eigenstate
fields as
\bea
 && \left(\begin{array}{c}
  H_2^0 \\
  H_1^0 \\
  \end{array}\right)
  =\left(\begin{array}{c}
  v_2\\
  v_1\\
  \end{array}\right)
  +{1\over \sqrt{2}}R_\alpha 
   \left(\begin{array}{c}
     h^0 \\
     H^0 \\
     \end{array}\right)
  +{i\over \sqrt{2}} R_{\beta_0}
   \left(\begin{array}{c}
    G^0 \\
    A^0 \\
    \end{array}\right) , \label{h12h}
\\ 
 && \left(\begin{array}{c}
   H_2^+ \\
   H_1^{-*} \\
  \end{array}\right)
  =R_{\beta_\pm}
 \left(\begin{array}{c}
   G^+ \\
   H^+ \\
   \end{array}\right)   .
\eea
The rotation matrices are given by
\bea
  && R_\alpha=\left(\begin{array}{cc}
  \cos \alpha & \sin \alpha \\
  -\sin \alpha & \cos \alpha \\ 
  \end{array}\right) ,~~
 R_{\beta_0}=\left(\begin{array}{cc}
  \sin \beta_0 & \cos \beta_0 \\
  -\cos \beta_0 & \sin \beta_0 \\
 \end{array}\right) ,
\\ 
 && R_{\beta_\pm}=\left(\begin{array}{cc}
  \sin \beta_\pm & \cos \beta_\pm \\
  -\cos \beta_\pm & \sin \beta_\pm \\
 \end{array}\right) .
\eea
The electrically-neutral part of the potential is
\bea
  V^{\textrm{\scriptsize neutral}}
  ={1\over 2} m_{h^0}^2 {h^0}^2
   +{1\over 2} m_{H^0}^2 {H^0}^2
   +{1\over 2} m_{G^0}^2 {G^0}^2
   +{1\over 2} m_{A^0}^2 {A^0}^2 ,
\eea
with
$\textrm{diag}(m_{h^0}^2, m_{H^0}^2)
=R^{-1}_\alpha m_{\phi_R^0}^2 R_\alpha$,
$\textrm{diag}(m_{G^0}^2, m_{A^0}^2)
=R^{-1}_{\beta_0} m_{\phi_I^0}^2 R_{\beta_0}$.
Here
\bea
  m_{\phi_R^0}^2&\!\!\!=\!\!\!& \left(\begin{array}{cc}
   |\mu|^2 +m_2^2 +{g^2+{g'}^2\over 4}(3v_2^2- v_1^2)
   & -b -{g^2 +{g'}^2\over 2} v_1 v_2 \\
   -b -{g^2 +{g'}^2\over 2} v_1 v_2
  & |\mu|^2 +m_1^2 +{g^2+{g'}^2\over 4}(3v_1^2- v_2^2)
   \end{array}
    \right) ,
    \label{eneure}
\\
 m_{\phi_I^0}^2 &\!\!\!=\!\!\!&
 \left(\begin{array}{cc}
   |\mu|^2 +m_2^2 +{g^2+{g'}^2\over 4}(v_2^2- v_1^2)
   & b  \\
   b 
  & |\mu|^2 +m_1^2 +{g^2+{g'}^2\over 4}(v_1^2- v_2^2)
   \end{array}\right) .
\eea
The electrically-charged part of the potential is
\bea
   V^{\textrm{\scriptsize charged}}
    =m_{G^\pm}^2 G^+ G^- +m_{H^\pm}^2 H^+ H^- ,
\eea
with
$\textrm{diag}(m_{G^\pm}^2, m_{H^\pm}^2)
=R^{-1}_{\beta_\pm} m_{\phi^\pm}^2 R_{\beta_\pm}$. Here
\bea
 m_{\phi^\pm}^2 =
 \left(\begin{array}{cc}
  |\mu|^2 +m_2^2 + {g^2+{g'}^2\over 4} v_2^2
   +{g^2-{g'}^2\over 4}  v_1^2
  & b+{g^2\over 2} v_1 v_2 \\
  b+{g^2\over 2} v_1 v_2  &
   |\mu|^2 +m_1^2 + {g^2+{g'}^2\over 4} v_1^2
   +{g^2-{g'}^2\over 4}  v_2^2 \\
 \end{array} \right).
\eea     
From the relation between the mass matrix $m_{\phi_R^0}^2$ and the eigenvalues, 
the following equations
are obtained:
\bea
  m_{A^0}^2+m_Z^2 &\!\!\!=\!\!\!& m_{h^0}^2 +m_{H^0}^2 ,
\\
  -{\sin 2\beta \over 2}m_{A^0}^2
 -{m_Z^2\over 2}\sin 2\beta &\!\!\!=\!\!\!& 
 (m_{H^0}^2 - m_{h^0}^2) {\sin 2\alpha\over 2} .
\eea
A formula is obtained as
\bea
  {\sin 2\alpha\over \sin 2\beta}
    =-{m_{A^0}^2 +m_Z^2\over m_{H^0}^2 -m_{h^0}^2}
     =-{m_{H^0}^2 +m_{h^0}^2\over m_{H^0}^2 -m_{h^0}^2} .
  \label{formula1}
\eea    
For $\beta_0=\beta$, the mass matrix $m_{\phi_I^0}^2$ 
leads to the mass eigenvalues
\bea
  m_{A^0}^2 ={2b\over \sin 2\beta} ,\quad
  m_{G^0}^2 =0.
  \label{ma0}
\eea
From these equations, 
the masses for $h^0$ and $H^0$ are derived,
\bea
   m_{h^0}^2 &\!\!\! =\!\!\!&
   {1\over 2}\left[
     m_{A^0}^2 +m_Z^2
     -\sqrt{(m_{A^0}^2 +m_Z^2)^2
     -4m_{A^0}^2 m_Z^2 \cos^2 2\beta}\right] ,
\\
   m_{H^0}^2 &\!\!\! =\!\!\!&
   {1\over 2}\left[
     m_{A^0}^2 +m_Z^2
     +\sqrt{(m_{A^0}^2 -m_Z^2)^2
     +4m_{A^0}^2 m_Z^2 \sin^2 2\beta}\right] .
\eea
For electrically-charged scalars, the mass eigenvalues are
\bea
  m_{H^\pm}^2 = m_{A^0}^2 +m_W^2 ,\quad
  m_{G^\pm}^2=0 .
\eea
The lightest Higgs boson is $h^0$.
The mass has the upper limit
\bea
   m_{h^0}^2 \leq m_{h^0}^2(m_{A^0}\to\infty)= m_Z^2 \cos^2 2\theta ,
\eea
at tree level. The tree-level value is experimentally excluded.

\section{Radiative corrections in an effective
four-dimensional picture
\label{rad}}

In this section we study the mass of the lightest Higgs boson
at loop level.
We start with review for radiative corrections
to the mass of the lightest Higgs boson in four-dimensional case~%
\cite{Martin:1997ns}\cite{Weinberg}.
Top quark and its superpartner, stop largely contribute to Higgs masses 
through radiative corrections.
We concentrate on top and stop contributions.
The MSSM superpotential includes
\bea
   W_t=y_t \bar{t}t H_2^0 -\mu H_2^0 H_1^0 .
   \label{wt}
\eea
where $y_t$ is the Yukawa coupling of the top quark $t$.
For the left-handed top $t_L$ and the right-handed top $\bar{t}_R$,
there are stop $\tilde{t}_L$ and $\tilde{\bar{t}}_R$.
The supersymmetry-breaking Lagrangian relevant to the stop masses is
\bea
   {\cal L}_{\textrm{\scriptsize soft}}
   =-\left[\tilde{\bar{t}}_R a_2 \tilde{t}_L H_2^0
    +\textrm{H.c.}\right]
   -\tilde{t}_L^\dag m_Q^2 \tilde{t}_L
  -\tilde{\bar{t}}_R m_{\bar{t}}^2 \tilde{\bar{t}}_R^\dag .
 \label{lsoft}
\eea
where supersymmetry-breaking coupling constants are denoted as
$a_2$, $m_Q$, $m_{\bar{t}}$.
From Eqs.(\ref{wt}) and (\ref{lsoft}), the stop mass terms are
\bea
    (\tilde{t}_L ~ {\tilde{\bar{t}}}_R^*)
  \left(\begin{array}{cc}
      m_Q^2 +|y_t H_2^0|^2 & a_2 H_2^0 -y_t \mu^* {H_1^0}^* 
 \\
    a_2^* {H_2^0}^* -y_t^* \mu {H_1^0}^* &
     m_{\bar{t}}^2 + |y_t H_2^0|^2 
  \end{array}\right)
  \left(\begin{array}{c}
     \tilde{t}_L^* \\
       {\tilde{\bar{t}}}_R 
   \end{array}\right) .
\eea
which are written as field-dependent mass terms.

For simplicity, 
we assume the field-dependent mass squared for
the two stops $\tilde{t}_L$ and $\tilde{\bar{t}}_R$ is
$m_{\tilde{t}}^2 +|y_t H_2^0|^2$.
Then the potential is radiatively corrected by
the sum of stop and top contributions given in
\bea
  {3\over (4\pi)^2}(m_{\tilde{t}}^2 + |y_t H_2^0|^2)^2
    \left(\ln {m_{\tilde{t}}^2 + |y_t H_2^0|^2\over 
  \Lambda^2} -{1\over 2}\right)
  -{3\over (4\pi)^2}|y_t H_2^0|^4 \left(
  \log {|y_t H_2^0|^2\over \Lambda^2} -{1\over 2}\right) .
    \label{effp}
\eea
With an expansion by powers of $|y_t H_2|^2/m_{\tilde{t}}^2$,
the potential is
\bea
   -{3\over 16\pi^2}|y_t H_2^0|^4
    \left(\ln {|y_t H_2^0|^2\over m_{\tilde{t}}^2 
  + |y_t H_2^0|^2}
   -{3\over 2}\right) \equiv U_t(|H_2^0|^2),
    \label{pou}
\eea
up to constant ($H_2^0$-independent) terms, higher order terms
and linear $|H_2^0|^2$ terms which are adjusted so that the stationary condition
is satisfied.
The potential (\ref{pou}) is written around $(\textrm{Re}\,H_2^0)^2=v_2^2$ as
\bea
 U_t((\textrm{Re}\,H_2^0)^2) =U_t(v_2^2)
  +((\textrm{Re}\,H_2^0)^2-v_2^2) U'_t(v_2^2)
  +{1\over 2}((\textrm{Re}\,H_2^0)^2-v_2^2)^2 U''_t(v_2^2) +\cdots .
\eea
By this contribution,
the mass matrix (\ref{eneure}) of the electrically-neutral real fields  
is corrected to
\bea
  m_{\phi_R^0}^2&\!\!\!=\!\!\!& \left(\begin{array}{cc}
   |\mu|^2 +m_2^2 +{g^2+{g'}^2\over 4}(3v_2^2- v_1^2) +\Delta_t
   & -b -{g^2 +{g'}^2\over 2} v_1 v_2 \\
   -b -{g^2 +{g'}^2\over 2} v_1 v_2
  & |\mu|^2 +m_1^2 +{g^2+{g'}^2\over 4}(3v_1^2- v_2^2)
   \end{array}
    \right) .
\eea
Here the top-stop correction is
\bea
  \Delta_t &\!\!\!=\!\!\!&
    2v_2^2 U''_t (v_2^2)
  =2v_2^2
    \left(-{3\over 8\pi^2}|y_t|^4
      \ln {|y_t|^2 v_2^2\over m_{\tilde{t}}^2}\right)
 ={3\sqrt{2}m_t^4 G_F\over 2\pi^2 \sin^2\beta}
 \ln\left({m_{\tilde{t}}^2\over m_t^2}\right) , 
 \label{contz}
\eea
with top quark mass $m_t^2=|y_t|^2 v_2^2$ and Fermi coupling constant
$G_F={1\over 2\sqrt{2}}(v_1^2+v_2^2)^{-1}= 1.17\times 10^{-5}~\textrm{GeV}^{-2}$.    
Thus, in the present approximation the lightest Higgs mass is
\bea
    m_{h^0}^2 = 
 {1\over 2}\left[m_{A^0}^2 +m_Z^2
   +\Delta_t -\sqrt{((m_{A^0}^2-m_Z^2)\cos 2\beta +\Delta_t)^2 
    +(m_{A^0}^2 +m_Z^2)^2\sin^2 2\beta }\right] .    
 \label{lightest}
\eea  
The upper bound to the mass is given by
\bea
  {m_{h^0}}^2 &\!\!\! \leq \!\!\!& {m_{h^0}}^2 (m_A\to \infty)
  = m_Z^2 \cos^2 2\beta +\Delta_t \sin^2\beta
\nonumber
\\
  &\!\!\!=\!\!\!& m_Z^2 \cos^2 2\beta +{3\sqrt{2}m_t^4 G_F\over 2\pi^2}
 \ln\left({m_{\tilde{t}}^2\over m_t^2}\right) , 
\eea
where Eq.~(\ref{contz}) has been used.
The Higgs boson mass is significantly lifted to by the radiative correction
of top and stop.

\subsubsection*{Kaluza-Klein mode contributions}

Next we consider the case where top and stop propagate in bulk.
In a four-dimensional language, zero mode and Kaluza-Klein modes give
contributions.
The contribution to the potential from the $n$-th Kaluza-Klein states
of bulk top and stop is 
\bea
   &&{3\over (4\pi)^2}\left(m_{\tilde{t}}^2 +|y_t H_2|^2 +{n^2\over R^2}\right)^2
     \left(\ln {m_{\tilde{t}}^2 +|y_t H_2^0|^2 +{n^2/ R^2}
     \over \Lambda^2} -{1\over 2}\right)
\nonumber
\\
  && -{3\over (4\pi)^2}\left(|y_t H_2^0|^2 +{n^2\over R^2}\right)^2
     \left(\ln {|y_t H_2^0|^2 +n^2/R^2
     \over \Lambda^2} -{1\over 2}\right) ,
     \label{effpKK}
\eea
analogous to Eq.(\ref{effp}).
Here $m_{\tilde{t}}^2 > |y_t H_2^0|^2$.
For illustration, in this section we adopt
the $n$-th top Kaluza-Klein mass squared $n^2/R^2$ and 
the $n$-th stop Kaluza-Klein mass squared
$(m_{\tilde{t}}^2 +n^2/R^2)$.
In this case, we explicitly show that the sum over contributions from an infinite number of fields with mass splitting is finite. An importance of the sum over an inifinite number of Kaluza-Klein modes to utilize five-dimensional invariance has been discussed~%
\cite{Barbieri:2000vh,ArkaniHamed:2001mi,ArkaniHamed:2000xj}.
For logarithmic terms, we perform the expansion by powers of
$(m_{\tilde{t}}/(nR^{-1}))$.
For example, 
\bea
   &&\ln {m_{\tilde{t}}^2 + |y_t H_2^0|^2 
   +{n^2/R^2}\over \Lambda^2}
   =\ln\left({n^2\over R^2\Lambda^2}
    \left(1+{R^2\over n^2}(m_{\tilde{t}}^2+ |y_t H_2^0|^2)\right)\right)
\nonumber
\\
 &&=\ln {n^2\over R^2\Lambda^2}
   +{R^2\over n^2} (m_{\tilde{t}}^2+ |y_t H_2|^2) 
   -{R^4\over 2n^4} (m_{\tilde{t}}^2+ |y_t H_2|^2)^2 +\cdots .
\eea  
In this way, the coefficients of a logarithmic term in 
Eq.~(\ref{effpKK}),
$(m_{\tilde{t}}^2+ |y_t H_2^0|^2+ n^2/R^2)$ may be split into
$(m_{\tilde{t}}^2+|y_t H_2^0|^2)$ and
$(n^2/R^2)$ parts.
%
%
Then Eq.~(\ref{effpKK}) is lead to
\bea
  && {3\over 16\pi^2}m_{\tilde{t}}^2 (m_{\tilde{t}}^2 + 2 |y_t H_2^0|^2 
   +2{n^2\over R^2})
 \ln {n^2\over R^2\Lambda^2}
 +{3\over 16\pi^2}m_{\tilde{t}}^2(m_{\tilde{t}}^2 + 2 |y_t H_2^0|^2)
\nonumber
\\
  && +{R^2\over 16\pi^2 n^2}((m_{\tilde{t}}^2
  +|y_t H_2^0|^2)^3 - |y_t H_2^0|^6)
\nonumber
\\
 &\!\!\!=\!\!\!&
  {3\over 16\pi^2}m_{\tilde{t}}^2\left(m_{\tilde{t}}^2+{2n^2\over R^2}\right)
 \ln {n^2\over R^2\Lambda^2}
  +{3\over 16\pi^2}m_{\tilde{t}}^4\left(1+{R^2\over 3n^2}m^2\right)
\nonumber
\\
  &&+{3\over 8\pi^2}m_{\tilde{t}}^2 |y_t H_2^0|^2 
  \ln {n^2\over R^2\Lambda^2}
+ {3\over 16\pi^2}m_{\tilde{t}}^2 |y_t H_2^0|^2
  \left(2+{R^2\over n^2}m^2\right)
+{3\over 16\pi^2}{R^2\over n^2}m_{\tilde{t}}^2 |y_t H_2^0|^4 .
\nonumber
\\
 && \label{kkpts}
\eea 
Here the first two terms are constant.
In the last line, the first and second terms are both linear,
which are treated as in the four-dimensional case.
Note the equilibrium condition is not
independent of each $n$-th Kaluza-Klein mode.  
For the sum of every mode, $v_2$ is adjusted.
Thus 
the radiative correction of the potential relevant to the Higgs mass
is the last term in Eq.~(\ref{kkpts}),
a quartic field term.

Now we have obtained the $n$-th contribution for the potential
\bea
 U_t^{(n)}(|H_2^0|^2) = {3\over 16\pi^2}{R^2\over n^2}
   m_{\tilde{t}}^2 |y_t H_2^0|^4  .
\eea
From this equation, the correction for the lightest Higgs boson mass matrix is
\bea
  \Delta_t^{(n)}
  =2v_2^2 {U_t^{(n)}}'' (v_2^2)
  ={3\over 4\pi^2}{R^2\over n^2} m_{\tilde{t}}^2
   |y_t|^4 v_2^2
 ={3\sqrt{2}\over 2\pi^2} {m_t^4 G_F\over \sin^2\beta}
  {R^2\over n^2} m_{\tilde{t}}^2 .
 \label{corecn}
\eea 
The mass of the lightest Higgs boson (\ref{lightest})
is corrected by amount of Kaluza-Klein mode contributions to
\bea
    m_{h^0}^2 &\!\!\!=\!\!\!& 
 {1\over 2}\left[m_{A^0}^2 +m_Z^2
   +\Delta_t +\sum_{n=1}^\infty \Delta_t^{(n)}
    \right.
\nonumber
\\
&&\left.
   -\sqrt{\left((m_{A^0}^2-m_Z^2)\cos 2\beta +\Delta_t
  +\sum_{n=1}^\infty \Delta_t^{(n)}\right)^2 
    +(m_{A^0}^2 +m_Z^2)^2\sin^2 2\beta }\right] .    
     \label{higgsk}
\eea
Due to the $(1/n^2)$-dependence of the mass, 
the sum of an infinite number of contributions converges,
\bea
   && \sum_{n=1}^\infty {1\over n^2} 
    ={\pi^2\over 6} \simeq 1.6 , \quad
   \sum_{n=1}^{10} {1\over n^2}
    \simeq 1.5 .
     \label{zeta}
\eea
In order to show that small Kaluza-Klein numbers are dominant, 
we have also written down
the sum for the first 10 Kaluza-Klein states.
Such an importance of small numbers of Kaluza-Klein modes has been shown
also in the context of corrections to gauge coupling constants~%
\cite{Uekusa:2007im, Uekusa:2008iz}. 
From Eq.~(\ref{higgsk}),
the upper bound to the lightest Higgs mass is given by
\bea
  {m_{h^0}}^2 &\!\!\! \leq \!\!\!& {m_{h^0}}^2 (m_A\to \infty)
  = m_Z^2 \cos^2 2\beta 
   +\left(\Delta_t +\sum_{n=1}^\infty \Delta_t^{(n)}\right) \sin^2\beta
\nonumber
\\
  &\!\!\!=\!\!\!& m_Z^2 \cos^2 2\beta +{3\sqrt{2}m_t^4 G_F\over 2\pi^2}
 \ln\left({m_{\tilde{t}}^2\over m_t^2}\right)
  +{\sqrt{2}\over 4} m_t^4 G_F (Rm_{\tilde{t}})^2  .
 \label{mhlbk}
\eea
In the last equation,
Eqs.~(\ref{contz}), (\ref{corecn}) and (\ref{zeta}) have also been used.

\section{The mass-eigenvalue splitting with bulk and brane 
supersymmetry breaking
\label{SS}}

We derive stop mass in an orbifold model with 
the Standard Model gauge bosons, third generation of quarks and leptons and 
two Higgs fields and their superpartners as five-dimensional bulk fields,
in similar to the model given in Refs.~\cite{DiClemente:2001sv,DiClemente:2002qa}.
And then we obtain correction to the Higgs boson mass.
Here zero modes are the MSSM fields. 
Bulk fields are classified into gauge multiplet and hypermultiplet.
The components of a gauge multiplet are a gauge field $A_M$, 
a adjoint scalar $\Sigma$ and two gaugino $\tilde{A}^i$,
where $M,N=0,1,2,3,5$ and $m,n=0,1,2,3$.
The gaugino $\tilde{A}^i$ is SU(2)${}_R$ doublet ($i=1,2$). 
The component of a hypermultiplet are SU(2)${}_R$-doublet scalars $\tilde{\psi}^i$
and a Dirac fermion $\psi$.
Fields which propagate in bulk give Kaluza-Klein contributions
in addition to zero mode contributions.

In an off-shell formulation, the gauge multiplet contains
three adjoint auxiliary fields $X^a$, which form a triplet of SU(2)${}_R$.
The five-dimensional Lagrangian for the gauge multiplet is given by 
\cite{Mirabelli:1997aj, ArkaniHamed:2001tb, Hebecker:2001ke}
\bea
{\cal L}_{\textrm{\scriptsize gauge}}={1\over g^2} \textrm{tr}
   \left(-\textrm{${1\over 2}$} (F_{MN})^2
  -(D_M \Sigma)^2
  -\bar{\tilde{A}}_ii\gamma^M D_M \tilde{A}^i
 + (X^a)^2
 +\bar{\tilde{A}}_i \textrm{[}\Sigma,\tilde{A}^i\textrm{]}
 \right) ,
\eea
with the supersymmetry transformation law
\bea
  \delta_\xi A^M &\!\!\!=\!\!\!& i\bar{\xi}_i\gamma^M\tilde{A}^i ,
\\
  \delta_\xi \Sigma &\!\!\!=\!\!\!& i\bar{\xi}_i\tilde{A}^i ,
\\
   \delta_\xi \tilde{A}^i &\!\!\!=\!\!\!& 
  \left(\textrm{${1\over 2}$}\gamma^{MN}F_{MN}+\gamma^M D_M\Sigma
  \right) \xi^i
  +i\left(X^a\sigma^a\right)^i_{\cdot j}
  \xi^j ,
\\
   \delta_\xi X^a &\!\!\!=\!\!\!& 
   \bar{\xi}_i(\sigma^a)^i_{\cdot j}\gamma^M D_M\tilde{A}^j
 +i\textrm{[}\Sigma,\bar{\xi}_i(\sigma^a)^i_{\cdot j}
  \tilde{A}^j\textrm{]} .
\eea
Here Dirac matrices are
\bea
\gamma^M=\left(\,\left(\begin{array}{cc}0&\sigma^m\\ \bar{\sigma}^m&0
\end{array}\right),\left(\begin{array}{cc}-i&0\\ 0&i\end{array}\right)\,
\right) ,
\eea
with $\sigma^m=(1,\vec{\sigma})$, $\bar{\sigma}^m=(1,-\vec{\sigma})$, 
and $\gamma^{MN}={1\over 2}\textrm{[}\gamma^M,\gamma^N\textrm{]}$.
The covariant derivative acts on fields like
$D_M \Sigma= \partial_M \Sigma +i[A_M , \Sigma]$. 
The symplectic Majorana condition is
\bea
\tilde{A}^i=\epsilon^{ij}C\bar{\tilde{A}}_j^T\,,
\eea
where the five-dimensional charge conjugation matrix $C$ satisfies 
$C\gamma^MC^{-1}=(\gamma^M)^T$. 
The explicit form is $C=\textrm{diag}(i\sigma^2,i\sigma^2)$. 
Lower indices $i,j,..$ transform under the $2^*$ of SU(2)${}_R$. 
The $\epsilon$ tensor is employed to
raise or lower indices where $\epsilon^{12}=\epsilon_{21}=1$.
The two four-dimensional Weyl spinors $\tilde{A}_L$ and $\tilde{A}_R$ are 
incorporated in
the expression of symplectic Majorana spinors as
\bea
\tilde{A}^1=
  \left(\begin{array}{cc}
   (\tilde{A}_L)_\alpha \\ 
   (\bar{\tilde{A}}_R)^{\dot{\alpha}} \\
   \end{array}\right) ,
   ~~
  \tilde{A}^2=
  \left(\begin{array}{cc}
  (\tilde{A}_R)_\alpha \\ 
 -(\bar{\tilde{A}}_L)^{\dot{\alpha}} \\
  \end{array}\right) ,
  ~~
  \bar{\tilde{A}}_1=
 \left(\begin{array}{cc}
  (\tilde{A}_R)^\alpha \\ 
  (\bar{\tilde{A}}_L)_{\dot{\alpha}} \\
  \end{array}\right)^T ,
  ~~
  \bar{\tilde{A}}_2=
  \left(\begin{array}{cc}
   -(\tilde{A}_L)^\alpha \\ 
   (\bar{\tilde{A}}_R)_{\dot{\alpha}} \\
  \end{array}\right)^T .
\eea
The supersymmetry transformation parameter 
is a symplectic Majorana spinor $\xi^i$.

The hypermultiplet contains
an SU(2)${}_R$-doublet auxiliary field $F_i$ in an off-shell formulation.
The five-dimensional gauge-interacting Lagrangian for the hypermultiplet is
\bea
{\cal L}_{\textrm{\scriptsize hyper}} 
  &\!\!\!=\!\!\!& -(D_M\tilde{\psi})^\dag_i(D^M\tilde{\psi}^i)
  -i\bar{\psi}\gamma^M D_M \psi
  +F^{\dag i}F_i-\bar{\psi} \Sigma \psi
  +\tilde{\psi}^\dag_i(\sigma^aX^a)^i_{\cdot j} \tilde{\psi}^j
\nonumber
\\
  && +\tilde{\psi}^\dag_i\Sigma^2 \tilde{\psi}^i
  +\left(i\sqrt{2}\bar{\psi}\tilde{A}^i\epsilon_{ij}
   \tilde{\psi}^j+\textrm{H.c.} 
  \right) ,
\eea
with the supersymmetry transformation law
\bea
  \delta_\xi \tilde{\psi}^i &\!\!\!=\!\!\!& 
  -\sqrt{2}\epsilon^{ij}\bar{\xi}_j\psi ,
\\
 \delta_\xi \psi &\!\!\!=\!\!\!& 
  i\sqrt{2}\gamma^M D_M \tilde{\psi}^i\epsilon_{ij}\xi^j
  -\sqrt{2}\Sigma
  \tilde{\psi}^i\epsilon_{ij}\xi^j
   +\sqrt{2}F_i\xi^i ,
\\
  \delta_\xi F_i &\!\!\!=\!\!\!& 
   i\sqrt{2}\bar{\xi}_i\gamma^M D_M \psi
   +\sqrt{2}\bar{\xi}_i \Sigma\psi
   -2i\bar{\xi}_i \tilde{A}^j\epsilon_{jk}\tilde{\psi}^k .
\eea

\subsubsection*{Parity and twist}

For the space of the extra-dimensional coordinate $y$, there are fixed points of
orbifold.
Fields must be consistently defined around the fixed points.
For the parity $Z_0:y\to -y$,
the boundary conditions for fields of the gauge and hypermultiplet are
\bea
&&A_\mu (x,-y)=P_0 A_\mu (x,y)P_0^\dag , ~~
 A_y (x,-y)=-P_0 A_y (x,y)P_0^\dag ,
\\
&&\Sigma (x,-y)=-P_0 \Sigma (x,y)P_0^\dag , ~~
 \tilde{A}^i (x,-y)=(\sigma^3)^i_{\cdot j} ~i\gamma_5 P_0 \tilde{A}^j (x,y)P_0^\dag ,
  \label{su21}
\\
&&
  \tilde{\psi}^i(x,-y)=(\sigma^3)^i_{\cdot j} \eta_0 P_0 \tilde{\psi}^j(x,y) , ~~
 \psi(x,-y) =\eta_0 P_0 i\gamma_5 \psi(x,y) ,
 \label{su22}
\eea
where  $P_0^2={\bf 1}$ and a sign parity is denoted as $\eta_0$. 
The fermions and SU(2)${}_R$-doublet scalar are written with their components 
as
\bea
  \left(\begin{array}{c}
  \tilde{A}_L   \\
  \tilde{A}_R   \\ 
   \end{array}\right)(x,-y)
&\!\!\!=\!\!\!& \left(\begin{array}{c}
 P_0\tilde{A}_L(x,y)P_0^\dag \\ 
 -P_0\tilde{A}_R(x,y)P_0^\dag \\
 \end{array}\right) , 
 ~~
  \left(\begin{array}{c}
    \tilde{\psi}^1 \\
    \tilde{\psi}^2 \\
  \end{array}\right)(x,-y)
  =\left(\begin{array}{c}
     \eta_0 P_0 \tilde{\psi}^1(y) \\
    -\eta_0 P_0 \tilde{\psi}^2(y) \\
     \end{array}
  \right) ,
\\
  \psi(-y)&\!\!\!=\!\!\!&\left(\begin{array}{c}
     \psi_L \\
     \psi_R \\
     \end{array}\right)(-y)
 =\left(\begin{array}{c}
  \eta_0 P_0 \psi_L(y) \\
 -\eta_0 P_0 \psi_R(y) \\
     \end{array}\right) .
\eea
For the parity $S:y\to y+2\pi R$, the boundary conditions for
SU(2)${}_R$-singlet bosonic fields are given by
\bea
 &&A_\mu (x,y+2\pi R) =V A_\mu (x,y)V^\dag , ~~
 A_y (x,y+2\pi R) =V A_y (x,y)V^\dag ,
\\
 &&\Sigma (x,y+2\pi R) =V \Sigma (x,y)V^\dag ,
\eea
where $V$ is a representation for $S$.
With the other parity $Z_1:\pi R+y\to \pi R-y$, the consistency conditions
$Z_0^2=Z_1^2=I$, $S=Z_1Z_0$ and $SZ_0S=Z_0$ must be fulfilled, where
$I$ is the identity operator.
As long as these conditions are satisfied,
$S$ has degrees of freedom of twisting
\bea
    U=\exp\left(i\sum_{j=1}^3 \beta_j \sigma_j\right)
    = \cos \beta \cdot {\bf 1}
      + {\sin\beta\over \beta} \sum_{j=1}^3 \beta_j \sigma_j ,
\eea
where $\beta=\sqrt{\beta_1^2+\beta_2^2+\beta_3^2}$.
For the $\sigma_3$-rotation of SU(2)${}_R$ around $Z_0$ given in Eqs.~(\ref{su21})
 and (\ref{su22}), the consistency conditions lead to
\bea
   \beta_3=0 ~~\textrm{or} ~~
   \beta_1=\beta_2=0 .
\eea
Then the representation of a twist about $S$ is
\bea
 U  &\!\!\!=\!\!\!&
  \left(\begin{array}{cc}
    \cos \beta & i (\beta_1-i\beta_2)\beta^{-1} \sin \beta \\
    i(\beta_1+i\beta_2)\beta^{-1}\sin\beta & \cos \beta \\
    \end{array}\right) ~~\textrm{for}~~ \beta_3=0 ;
\nonumber
\\
 &\!\!\!=\!\!\!&
  \left(\begin{array}{cc}
    \cos\beta &\sin\beta \\
    -\sin\beta & \cos \beta
    \end{array}\right) ~~\textrm{for}~~ \beta_3=\beta_1=0 .
    \label{twistu}
\eea
Hereafter we assume $\beta_3=\beta_1=0$.
The remaining fields has the parity for $S$ given by
\bea
 \tilde{A}^i(x,y+2\pi R)&\!\!\!=\!\!\!&
   \left(\begin{array}{cc}
     \cos\beta & \sin \beta \\
     -\sin\beta &\cos \beta 
     \end{array}\right)^i_{\cdot j}
   V\tilde{A}^j(x,y)V^\dag  ,
\\   
  \tilde{\psi}^i(y+2\pi R) &\!\!\!=\!\!\!& \left(\begin{array}{cc}
    \cos\beta & \sin\beta \\
    -\sin\beta & \cos\beta
    \end{array}\right)^i_{\cdot j}
    \eta_0\eta_1 V \tilde{\psi}^j(y) ,
\\
  \psi(y+2\pi R)&\!\!\!=\!\!\!& \eta_0\eta_1 V\psi(y) ,
\eea   
where $\eta_1$ is a sign parity.
In components, for example, the gaugino is written as
\bea
 \left(\begin{array}{c}
   \tilde{A}_L \\
   \tilde{A}_R \\
   \end{array}\right)
   (x,y+2\pi R)=
   \left(\begin{array}{cc}
    \cos\beta & \sin\beta \\
    -\sin\beta & \cos\beta
    \end{array}\right)
    \left(\begin{array}{c}
     V\tilde{A}_L(x,y)V^\dag \\
     V\tilde{A}_R(x,y)V^\dag
     \end{array}\right) .
\eea
The parity for $Z_1$ is read from the parities for $Z_0$ and $S$ via $V=P_1 P_0$.
The boundary conditions for SU(2)${}_R$-singlet fields are
\bea
 A_\mu(x,\pi R-y)&\!\!\!=\!\!\!& P_1 A_\mu(x,\pi R+y)P_1^\dag , ~~
  A_y(x,\pi R-y)=-P_1 A_y(x,\pi R+y)P_1^\dag ,
 \\
 \Sigma(x,\pi R-y)&\!\!\!=\!\!\!&-P_1 \Sigma(x,\pi R+y)P_1^\dag ,~~
 \psi(x,\pi R-y)=\eta_1 P_1 i\gamma_5 \psi(x,\pi R+y) ,
\eea
and the boundary conditions for SU(2)${}_R$-doublet fields are
\bea
  \left(\begin{array}{c}
   \tilde{A}_L \\
   \tilde{A}_R 
   \end{array}
   \right) 
    (x,\pi R-y) &\!\!\!=\!\!\!&
    \left(\begin{array}{cc}
    \cos\beta & \sin\beta \\
    -\sin\beta & \cos\beta
    \end{array}\right)
     \left(\begin{array}{c}
     P_1\tilde{A}_L(x,\pi R+y)P_1^\dag \\
     -P_1\tilde{A}_R(x,\pi R+y)P_1^\dag
     \end{array}\right) ,
\\
  \left(\begin{array}{c}
    \tilde{\psi}^1\\
    \tilde{\psi}^2
    \end{array}
    \right)(x,\pi R-y)
    &\!\!\!=\!\!\!& 
 \left(\begin{array}{cc}
    \cos\beta & \sin\beta \\
    -\sin\beta & \cos\beta
    \end{array}\right)
    \left(\begin{array}{c}
    \eta_1P_1 \tilde{\psi}^1(x,\pi R+y) \\
    -\eta_1 P_1 \tilde{\psi}^2 (x,\pi R+y) \\
    \end{array}\right) .
\eea

\subsubsection*{Mode functions}

From the equations of the parities given above, we write down 
mode functions.
Let us consider the case where $P_0=P_1=V=1$, $\eta_0=\eta_1=1$.
Other cases with various signs are obtained by appropriate
exchanges of cosines and sines appearing in the following.
The gauge and hypermultiplet fields are mode-expanded as
\bea
 && A_\mu(x,y) =\sqrt{1\over \pi R} A_\mu^0(x)
   +\sqrt{2\over \pi R} A_\mu^n(x) \cos {ny\over R} ,
 \\
 && A_y(x,y) =\sqrt{2\over \pi R} A_y^n (x)\sin {ny\over R}, \quad
   \Sigma (x,y)=\sqrt{2\over \pi R} \Sigma^n (x)\sin {ny\over R} ,
 \\
 &&\left(\begin{array}{c}
 \tilde{A}_L\\
 \tilde{A}_R
 \end{array}\right)(x,y)
  =\textrm{$\sqrt{2\over \pi R}$}
   \left(\begin{array}{cc}
     \cos {\beta y\over 2\pi R} & \sin {\beta y\over 2\pi R} \\
     -\sin {\beta y\over 2\pi R} & \cos {\beta y\over 2\pi R}
    \end{array}\right)
    \left(\begin{array}{c}
 {1\over \sqrt{2}}\tilde{A}^{1,0}(x) +\tilde{A}^{1,n}(x)\cos {ny \over R} \\
 \tilde{A}^{2,n}(x)\sin{ny\over R}
 \end{array}\right) ,
\\ 
 &&\left(\begin{array}{c}
 \tilde{\psi}^1\\
 \tilde{\psi}^2
 \end{array}\right)(x,y)
  =\textrm{$\sqrt{2\over \pi R}$}
   \left(\begin{array}{cc}
     \cos {\beta y\over 2\pi R} & \sin {\beta y\over 2\pi R} \\
     -\sin {\beta y\over 2\pi R} & \cos {\beta y\over 2\pi R}
    \end{array}\right)
    \left(\begin{array}{c}
 {1\over \sqrt{2}}\tilde{\psi}^{1,0}(x) +\tilde{\psi}^{1,n}(x)\cos {n y\over R} \\
 \tilde{\psi}^{2,n}(x)\sin{n y\over R}
 \end{array}\right) ,
  \label{tilpsi}
\\ 
 &&\left(\begin{array}{c}
 \psi_L\\
 \psi_R
 \end{array}\right)(x,y)
  = \textrm{$\sqrt{2\over \pi R}$}\left(\begin{array}{c}
 {1\over \sqrt{2}}\psi^{1,0}(x) +\psi^{1,n}(x)\cos {ny\over R} \\
 \psi^{2,n}(x)\sin{ny\over R}
 \end{array}\right) .
\eea
For each contracted $n$, the summation $\sum_{n=1}^\infty$ has been 
taken.
Here the numerical factors are chosen from the normalization
\bea
  {1\over 2}\int_{-\pi R}^{\pi R} dy \,
   \sqrt{2\over \pi R} \cos {ny\over R}
   \cdot \sqrt{2\over \pi R} \cos {my\over R}
   =\delta_{nm} .
\eea
From the mode expansion give above,
an SU(2)${}_R$-doublet hyperscalar yield the mass term
\bea
&& {1\over 2}\int_{-\pi R}^{\pi R} dy
 \left(\begin{array}{c}
  \partial_y \tilde{\psi}^1 \\
  \partial_y \tilde{\psi}^2 
  \end{array}\right)^\dag
  \left(\begin{array}{c}
  \partial_y \tilde{\psi}^1 \\
  \partial_y \tilde{\psi}^2 
  \end{array}\right)
\nonumber
\\
  &\!\!\!=\!\!\!&\left({\beta\over 2\pi R}\right)^2 |\tilde{\psi}^{1,0}|^2
    +\left(\begin{array}{c}
      \tilde{\psi}^{1,n} \\
      \tilde{\psi}^{2,n} \\
      \end{array}\right)^\dag
      \left(\begin{array}{cc}
      \left({\beta\over 2\pi R}\right)^2 +{n^2\over R^2}
      & -2 \left({\beta\over 2\pi R}\right) {n\over R} \\
       -2 \left({\beta\over 2\pi R}\right) {n\over R} &
      \left({\beta\over 2\pi R}\right)^2 +{n^2\over R^2} \\
      \end{array}\right)
      \left(\begin{array}{c}
      \tilde{\psi}^{1,n} \\
      \tilde{\psi}^{2,n} \\
      \end{array} \right) .
  \label{bulkkin}
\eea
The mass matrix is diagonal with respect to Kaluza-Klein modes.
The $n$-th Kaluza-Klein mode has the squared-mass eigenvalue 
$(n\pm \beta/(2\pi))^2/R^2$.

\subsubsection*{Supersymmetry and its breaking}

The supersymmetry transformation parameter
is a symplectic Majorana spinor and is mode-expanded similarly to 
the gaugino $\tilde{A}^i$ as
\bea
  \left(\begin{array}{c}
 \xi_L\\
 \xi_R
 \end{array}\right)(x,y)
  =\textrm{$\sqrt{2\over \pi R}$}
   \left(\begin{array}{cc}
     \cos {\beta y\over 2\pi R} & \sin {\beta y\over 2\pi R} \\
     -\sin {\beta y\over 2\pi R} & \cos {\beta y\over 2\pi R}
    \end{array}\right)
    \left(\begin{array}{c}
 {1\over \sqrt{2}}\xi^{1,0}(x) +\xi^{1,n}(x)\cos {ny \over R} \\
 \xi^{2,n}(x)\sin{ny\over R}
 \end{array}\right) .
\eea
The zero mode corresponds to unbroken supersymmetry in four dimensions.
At $y=0$, the zero mode has the value
\bea
   \left(\begin{array}{c}
   \xi_L(x,y=0)\\
   \xi_R(x,y=0)\\
   \end{array}\right)_{\textrm{\scriptsize zero mode}}
    =\textrm{$\sqrt{1\over \pi R}$}
   \left(\begin{array}{cc}
     1 & 0 \\
     0 & 1
     \end{array}\right)
  \left(\begin{array}{c}
    \xi^{1,0}(x) \\
    0 \\
    \end{array}\right) .
\eea
There is $N=1$ supersymmetry at $y=0$ in the direction of $\xi_L$ because of
$\xi_R(y=0)=0$.
At $y=\pi R$, the zero mode has the value
\bea
  \left(\begin{array}{c}
 \xi_L(x,y=\pi R)\\
 \xi_R(x,y=\pi R)\\
 \end{array}\right)_{\textrm{\scriptsize zero mode}}
  =\textrm{$\sqrt{1\over \pi R}$}
   \left(\begin{array}{cc}
     \cos {\beta \over 2} & \sin {\beta \over 2} \\
     -\sin {\beta \over 2} & \cos {\beta \over 2} \\
    \end{array}\right)
    \left(\begin{array}{c}
 \xi^{1,0}(x)  \\
 0
 \end{array}\right) .
\eea
If a linear combination of $\xi_L$ and $\xi_R$ is taken as
\bea
 \left(\begin{array}{c}
 \xi_{\pi_1} \\
 \xi_{\pi_2} \\
 \end{array}\right)
  = \left(\begin{array}{cc}
     \cos {\beta \over 2} & -\sin {\beta \over 2} \\
     \sin {\beta \over 2} & \cos {\beta \over 2} \\
    \end{array}\right)
    \left(\begin{array}{c}
 \xi_L \\
 \xi_R \\
 \end{array}\right) ,
\eea 
there is $N=1$ supersymmetry at $y=\pi R$ in the direction of $\xi_{\pi_1}$
because of $\xi_{\pi_2}(y=\pi R)=0$.

At each boundary, four-dimensional supersymmetric couplings are possible 
for fields that do not vanish under multiplication of a delta function.
For $y=0$, nonzero fields are included in
$A_\mu$, $\tilde{A}_L$, $\tilde{\psi}^1$ and $\psi_L$.
For $y=\pi R$, nonzero fields are included in
$A_\mu$, $\tilde{A}_{\pi_1}$, $\tilde{\psi}_{\pi_1}$ and $\psi_L$.
Here linear combinations of SU(2)${}_R$-doublet fields are taken 
in the direction of $\xi_{\pi_1}$ for unbroken supersymmetry at $y=\pi R$ as
\bea
 \left(\begin{array}{c}
 \tilde{A}_{\pi_1} \\
 \tilde{A}_{\pi_2} \\
 \end{array}\right)
  = \left(\begin{array}{cc}
     \cos {\beta \over 2} & -\sin {\beta \over 2} \\
     \sin {\beta \over 2} & \cos {\beta \over 2} \\
    \end{array}\right)
    \left(\begin{array}{c}
 \tilde{A}_L \\
 \tilde{A}_R \\
 \end{array}\right) ,~~
 \left(\begin{array}{c}
 \tilde{\psi}_{\pi_1} \\
 \tilde{\psi}_{\pi_2} \\
 \end{array}\right)
  = \left(\begin{array}{cc}
     \cos {\beta \over 2} & -\sin {\beta \over 2} \\
     \sin {\beta \over 2} & \cos {\beta \over 2} \\
    \end{array}\right)
    \left(\begin{array}{c}
 \tilde{\psi}^1 \\
 \tilde{\psi}^2 \\
 \end{array}\right) .
  \label{rotationh}
\eea 
At $y=0$, $A_\mu(x,y)$ and $\tilde{A}_L(x,y)$ are formed into an $N=1$ gauge multiplet
and $\tilde{\psi}^1(x,y)$ and $\psi_L(x,y)$ are formed into a chiral multiplet.
For $y=\pi R$,
$A_\mu(x,y)$ and $\tilde{A}_{\pi_1}(x,y)$ are in a gauge multiplet
and $\tilde{\psi}_{\pi_1}(x,y)$ and $\psi_L(x,y)$ are in a chiral multiplet.
Because the directions of $\xi_L$ and $\xi_{\pi_1}$ are different from
each other, supersymmetry of total effective four-dimensional theory is
completely broken.

\subsubsection*{$F$-term on a brane}

Let us introduce $F$-term on the brane at $y=\pi R$. 
At $y=0$, the MSSM fields except for the bulk fields are confined.
We first review the case with no Scherk-Schwarz twist.
In this case, the mixing of Kaluza-Klein modes  and its diagonalization
was described in detail in \cite{DiClemente:2001sv}.
As in Section~\ref{rad},
we concentrate on top and stop contributions.
We denote a hypermultiplet with left-handed top quark $t_L$ 
in the zero-mode component as $T_{L}^i$ where $i=1,2$.
In this notation, the Lagrangian involving $T_{L}^i$ 
also allows supersymmetry transformation of the function $\xi^i$.
The stop $\tilde{t}_L^i$ is contained as scalar components in $T_L^i$.
The mode functions for $\tilde{t}_L^i$ are
given like the mode functions for $\tilde{\psi}^i$ in Eq.~(\ref{tilpsi}).
Similarly $T_{R}^i$ are defined for right-handed top quark $t_R$. 
The field on $y=\pi R$, $S$ 
is coupled to bulk matter superfields as
\bea
 {\cal L}_{\pi R}
 &\!\!\!=\!\!\!&\delta(y-\pi R) {1\over 2}\left[
   - {c_{l}\over \Lambda^3}T_{L}^{1^\dag} T_{L}^1
     S^\dag S
   -{c_r\over \Lambda^3}T_{R}^{1^\dag} T_{R}^1
     S^\dag S
   \right]_D ,
\eea
where we have used $T_{L}^2=T_{R}^2=0$ at $y=\pi R$ for no Scherk-Schwartz twist
and extracting $D$-term is represented as
the square brackets with the subscript $D$ such as $\left[~\right]_D$
with the notation of a vector superfield 
$V=C+{1\over 2}\theta\theta\bar{\theta}\bar{\theta}(D+{1\over 2}\Box C)
+\cdots$.
When the $S$ develops $\langle F_S\rangle$, the stop mass is generated.
With the dimensionless quantity
\bea
 \alpha=c_{\tilde{t}}\,\pi \left(
  {|\langle F_S\rangle |^2\over \Lambda^4}\right) \Lambda R ,
\eea
where we have assumed $c_l=c_r=c_{\tilde{t}}$ for simplicity,
the stop mass term is given by
\bea
 -{\cal L}_4^{\textrm{\scriptsize mass}}
  &\!\!\!=\!\!\!&{1\over R^2}\left[ \sum_{n=1}^\infty
  n^2 (\tilde{t}_{L,n}^{1*} \tilde{t}_{L,n}^1
    +\tilde{t}_{L,n}^{2*} \tilde{t}_{L,n}^2)
 +\sum_{n,m=1}^\infty {2\alpha\over \pi^2} (-1)^{n+m}
  \tilde{t}_{L,n}^{1*}\tilde{t}_{L,m}^1
 \right.
\nonumber
\\
  && \left. \qquad
   +\left(\sum_{n=1}^\infty {2\alpha\over \sqrt{2}\pi^2}(-1)^n
     \tilde{t}_{L,n}^{1*} \tilde{t}_{L,0}^1 + \textrm{H.c.}\right)
   +{\alpha\over \pi^2}\tilde{t}_{L,0}^{1*}\tilde{t}_{L,0}^1
    +(L\leftrightarrow R)\right] .
     \label{stma}
\eea
Eq.~(\ref{stma}) shows that the stop mass has the mixing between Kaluza-Klein modes.
If $\alpha\sim {\cal O}(1)$, mass eigenvalues are of the order $R^{-2}$.
For $R^{-1}\sim 1~\textrm{TeV}$, the zero-mode mass splitting
$m_{\tilde{t}}\sim R^{-1}$ lead to
significant radiative corrections to the Higgs boson mass.
In the upper bound (\ref{mhlbk}) to the lightest Higgs boson mass,
there are contributions largely from the second and third terms.
If $\alpha$ itself is very small and $m_{\tilde{t}}\sim \alpha R^{-1}$, the Higgs boson mass would be corrected
for a large $R^{-1}$ 
such as grand unification scale \cite{DiClemente:2002qa}.

\begin{figure}[htb]
\begin{center}
\includegraphics[width=8cm]{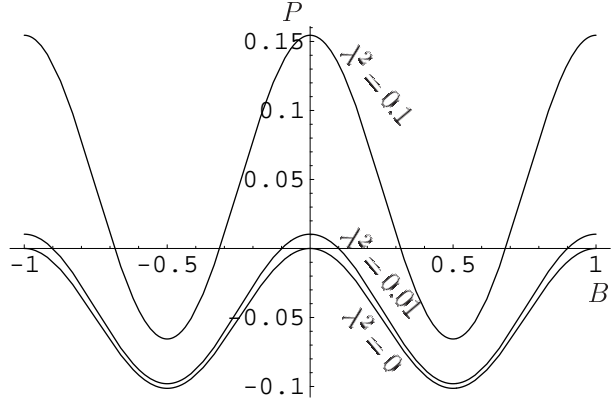} 
\caption{$P=C_{\tilde{t}}|\langle F_S\rangle|^2 
R \Lambda^{-3}/\pi$ and $B=\beta/(2\pi)$ for $\lambda^2=0$, 0.01, 0.1.
The summation over $n$ appearing in (\ref{gomasseq}) is taken as $\sum_{n=1}^{1000}$.
 \label{figa}}
\end{center}
\end{figure}

Now we consider supersymmetry breaking by the Scherk-Schwarz twist and 
$F$-term on a brane.
In the case with the Scherk-Schwarz twist (\ref{twistu}),
$N=1$ supersymmetry at $y=\pi R$ is in the direction of $\xi_{\pi_1}$.
Among the whole superfields, 
fields that do not vanish at $y=\pi R$ form 
$N=1$ superfields in the direction of $\xi_{\pi_1}$. 
For $T_{L}^i$, the bulk superfield oriented in the direction of
$\xi_{\pi_1}$ is denoted as
$T_{L\,\pi_1}$.  
The $\xi_{\pi_1}$-oriented superfield $T_{L\,\pi_1}$ contains
a stop $\tilde{t}_{L\,\pi_1}$ and a top $t_L^1$.
For $T_R^i$, the $\xi_{\pi_1}$-oriented $T_{R\,\pi_1}$ is defined in an analogous way. 
Here
\bea
\left(\begin{array}{c}
 \tilde{t}_{L\,\pi_1} \\
 \tilde{t}_{L\,\pi_2} \\
 \end{array}\right)
  = \left(\begin{array}{cc}
     \cos {\beta \over 2} & -\sin {\beta \over 2} \\
     \sin {\beta \over 2} & \cos {\beta \over 2} \\
    \end{array}\right)
    \left(\begin{array}{c}
 \tilde{t}_L^1 \\
 \tilde{t}_L^2 \\
 \end{array}\right) , ~~
\left(\begin{array}{c}
 \tilde{t}_{R\,\pi_1} \\
 \tilde{t}_{R\,\pi_2} \\
 \end{array}\right)
  = \left(\begin{array}{cc}
     \cos {\beta \over 2} & -\sin {\beta \over 2} \\
     \sin {\beta \over 2} & \cos {\beta \over 2} \\
    \end{array}\right)
    \left(\begin{array}{c}
 \tilde{t}_R^1 \\
 \tilde{t}_R^2 \\
 \end{array}\right) ,
\eea 
like Eq.~(\ref{rotationh}).
Then the boundary couplings are
\bea
 {\cal L}_{\pi R}
 &\!\!\!=\!\!\!&\delta(y-\pi R){1\over 2}\left[
   - {c_{l}\over \Lambda^3}T_{L\,\pi_1}^\dag T_{L\,\pi_1}
     S^\dag S
   - {c_{r}\over \Lambda^3}T_{R\,\pi_1}^\dag T_{R\,\pi_1}
     S^\dag S
   \right ]_D .
    \label{branestopss}
\eea
For nonzero twist and $F$-term,
we obtain the stop mass term
\bea
 -{\cal L}_4^{\textrm{\scriptsize mass}}
  &\!\!\!=\!\!\!&{1\over R^2}\left[
  {\beta^2\over 4\pi^2}\tilde{t}_{L,0}^{1*}\tilde{t}_{L,0}^1
  +\sum_{n=1}^\infty
  \left(n^2+{\beta^2\over 4\pi^2}\right)
   (\tilde{t}_{L,n}^{1*} \tilde{t}_{L,n}^1 +\tilde{t}_{L,n}^{2*} \tilde{t}_{L,n}^2)
  \right.
\nonumber
\\  
  &&-\sum_{n=1}^\infty {n\beta\over \pi}
 (\tilde{t}_{L,n}^{1*} \tilde{t}_{L,n}^2 +\tilde{t}_{L,n}^{2*} \tilde{t}_{L,n}^1)
 +\sum_{n,m=1}^\infty {2\alpha\over \pi^2} (-1)^{n+m}
  \tilde{t}_{L,n}^{1*}\tilde{t}_{L,m}^1
\nonumber
\\
  && \left. 
   +\sum_{k=n}^\infty {\sqrt{2}\alpha\over \pi^2}(-1)^n
    ( \tilde{t}_{L,n}^{1*} \tilde{t}_{L,0}^1 
  + \tilde{t}_{L,0}^{1*} \tilde{t}_{L,n}^1)
   +{\alpha\over \pi^2}\tilde{t}_{L,0}^{1*}\tilde{t}_{L,0}^1
    +(L\leftrightarrow R)\right] .
\eea
Define
\bea
   B={\beta\over 2\pi} , 
  \qquad
   P={\alpha\over \pi^2} .
\eea
Up to the overall $R^{-2}$,
$L$ part of mass term is written with a matrix form in a basis of
$(\tilde{t}_{L,0}^1,\tilde{t}_{L,1}^1,\tilde{t}_{L,2}^1,\cdots
|\tilde{t}_{L,1}^2,\tilde{t}_{L,2}^2,\cdots)^T$ as
\bea
 \left(\begin{array}{ccccc|cccc}
  B^2\textrm{+}P & -\sqrt{2}P & \sqrt{2}P & -\sqrt{2}P & \textrm{...} 
  & 0 & 0 & 0 & \textrm{...} \\
  -\sqrt{2}P & 1\textrm{+}B^2\textrm{+}2P & -2P & 2P & \textrm{...} 
  & -2B & 0 & 0 & \textrm{...} \\
  \sqrt{2}P & -2P & 4\textrm{+}B^2\textrm{+}2P & -2P & \textrm{...} 
  & 0 & -4B & 0 & \textrm{...} \\
  -\sqrt{2}P & 2P & -2P & 9\textrm{+}B^2 \textrm{+}2P &  \textrm{...} 
  & 0 & 0 & -6B & \textrm{...}\\ 
  &&&&\textrm{...} &&&&  \textrm{...}\\ \hline
  0 & -2B & 0 & 0 &  \textrm{...}
  &  1\textrm{+}B^2 & 0 & 0 &  \textrm{...}\\
  0 & 0 & -4B &0 & \textrm{...}
  & 0 & 4\textrm{+}B^2 & 0 &  \textrm{...}\\
  0 & 0& 0 & -6B &  \textrm{...}
  & 0 & 0 & 9\textrm{+}B^2  & \textrm{...}\\
  &&&&\textrm{...} &&&&  \textrm{...}\\
  \end{array}\right) .
  \label{mm}
\eea
In the first line of the upper-left block matrix, the ellipses
mean that $\pm\sqrt{2}P$ appear alternately.
In the second line, $\mp 2P$ appear alternately.
In order to diagonalize the mass matrix (\ref{mm}), we introduce
an eigenstate $(Q_0^1,Q_1^1,Q_2^1,\cdots|Q_1^2,Q_2^2,\cdots)^T$
and its eigenvalue $\lambda^2$.
Then eigenvalue equations  
are explicitly written as
\bea
 && (B^2+P)Q_0^1 -\sqrt{2}PS_o^1+\sqrt{2}PS_e^1 =\lambda^2 Q_0^1 ,
   \label{nzero}
\\  
 &&
-\sqrt{2}PQ_0^1+(n^2+B^2) Q_n^1+2PS_o^1 -2PS_e^1 -2nBQ_n^2 =\lambda^2 Q_n^1 ,
~~\textrm{for odd $n$},
 \label{nodd}
\\
&&
 \sqrt{2}PQ_0^1-2PS_o^1+(n^2+B^2) Q_n^1  +2PS_e -2nB Q_n^2 =\lambda^2 Q_n^1 ,
 ~~\textrm{for even $n$},
\\
 && -2nB Q_n^1 +(n^2+B^2) Q_n^2 =\lambda^2 Q_n^2 ,
  \label{q2}
\eea
where
$S_o^1 =Q_1^1 + Q_3^1+ Q_5^1 +\cdots$ and
$S_e^1 =Q_2^1 + Q_4^1 +Q_6^1 +\cdots$.
From Eq.~(\ref{q2}), $Q_n^2$ is solved by $Q_n^1$,
\bea
    Q_n^2 ={2nB\over n^2 +B^2 -\lambda^2} Q_n^1 .
\eea
With this equation, the equation for $n$ odd (\ref{nodd}) is
\bea
 {1\over 2}\left[ {1\over \lambda^2 -(n+B)^2}+{1\over \lambda^2 -(n-B)^2}\right]
   (-\sqrt{2}PQ_0^1 +2PS_o^1 -2P S_e^1) =Q_n^1 .
\eea
The summation with respect to odd $n$ gives
\bea
 {1\over \sqrt{2}}PQ_0^1 
 -\left(P-{1\over \Sigma_{oB}}\right)S_o^1 
 +PS_e^1 =0 ,
\eea
where 
\bea
 \Sigma_{oB}= \sum_{n ~\textrm{\scriptsize odd}}
    {1\over \lambda^2-(n+ B)^2} 
   +\sum_{n ~\textrm{\scriptsize odd}}
    {1\over \lambda^2-(n- B)^2} .
\eea
With a similar calculation, the equation for even $n$ lead to
\bea
  {1\over \sqrt{2}}PQ_0^1 -PS_0^1 
  +\left(P-{1\over \Sigma_{eB}}\right)S_e^1 =0 ,
\eea
where
\bea
   \Sigma_{eB} =\sum_{n~\textrm{\scriptsize even}}
       {1\over \lambda^2 -(n+ B)^2} 
  +\sum_{n~\textrm{\scriptsize even}}
       {1\over \lambda^2 -(n- B)^2} .
\eea
From these equations, the relation between $Q_0^1$ and $S_e^1$ is obtained as
\bea
  {P\over \sqrt{2}\Sigma_{oB}}Q_0^1
  +\left(P^2-\left(P-{1\over \Sigma_{oB}}\right)
   \left(P-{1\over \Sigma_{eB}}\right)\right)S_e^1 =0 .
  \label{rel1}
\eea
The relation between $Q_0^1$ and $S_o^1$ is 
\bea
  {P\over \sqrt{2}\Sigma_{eB}}Q_0^1
  +\left(\left(P-{1\over \Sigma_{eB}}\right)
   \left(P-{1\over \Sigma_{oB}}\right)-P^2\right)S_o^1 =0 .
   \label{rel2}
\eea
Substituting Eqs.~(\ref{rel1}) and (\ref{rel2}) into the equation for
$Q_0^1$ (\ref{nzero}), we obtain the eigenvalue equation
\bea
  {1\over \lambda^2-B^2}
    +\sum_{n=1}^\infty
      \left({1\over \lambda^2-(n+B)^2}
       +{1\over \lambda^2 -(n-B)^2}\right)
     ={1\over P} .
     \label{gomasseq}
\eea
Behaviors of $P$ and $B$
in Eq.(\ref{gomasseq}) are shown 
for several values of the eigenvalue $\lambda$ in Figure~\ref{figa}.
The eigenvalue equation (\ref{gomasseq}) is invariant under
the shift $B\to B'=B+1$. For the shift,
the left-hand side of (\ref{gomasseq}) is explicitly written as
\bea
{1\over \lambda^2-(B+1)^2}
    +\sum_{n=2}^\infty{1\over \lambda^2-(n+B)^2}
       +{1\over \lambda^2-B^2}
   +\sum_{n=1}^\infty{1\over \lambda^2 -(n-B)^2},
\eea
which is the same as the original.
From this point, 
invariance under the shift of $B$ with arbitrarily integer is induced.
The eigenvalue equation (\ref{gomasseq}) is also invariant under 
$B\leftrightarrow -B$.
Thus we deal with values of the twist for $0\leq B\leq 1/2$.
As seen in Figure~\ref{figa},
for $B=0$ (or $\beta=0$), a small eigenvalue is generated only for
a small $P\propto \alpha$. 
In other words, then the eigenvalue and $P$ are the same order.
When there is a nonzero Scherk-Schwarz twist, an interesting thing occurs.
For $B\sim 0.5$, an eigenvalue $\lambda^2\sim 0.01$ is
generated for $|P|\sim 0.1$ which is the same order as $B$.
Even if $|P|$ and $B$ are the same order and not very small,
a very small mass-eigenvalue can be generated due to
the mixing of effects of $P$ and $B$.
Taking the overall $R^{-2}$ back in Eq.~(\ref{mm}),
a TeV-scale mass splitting corresponds to
$R^{-1}\sim 1~\textrm{TeV}$ for $\lambda\sim 1$
and $R^{-1}\gg \textrm{TeV}$ for $\lambda\ll 1$. 

The appearance of small numbers with the mixing can be
expected if we concentrate on the upper-left block matrix of Eq.~(\ref{mm}).
In the upper-left block matrix, if $P=0$, $B$ and the eigenvalue of the block matrix
are the same order. If $B=0$, $P$ and the eigenvalue of the block matrix
are the same order. Then a very small eigenvalue corresponds to the value itself of
a very small $B$ or $P$. On the other hand,
if $P$ and $B$ are both nonzero and a large mixing occurs,
the eigenvalue can be much smaller than the values of $P$ and $B$.

\begin{figure}[htb]
\begin{center}
\includegraphics[width=6.5cm]{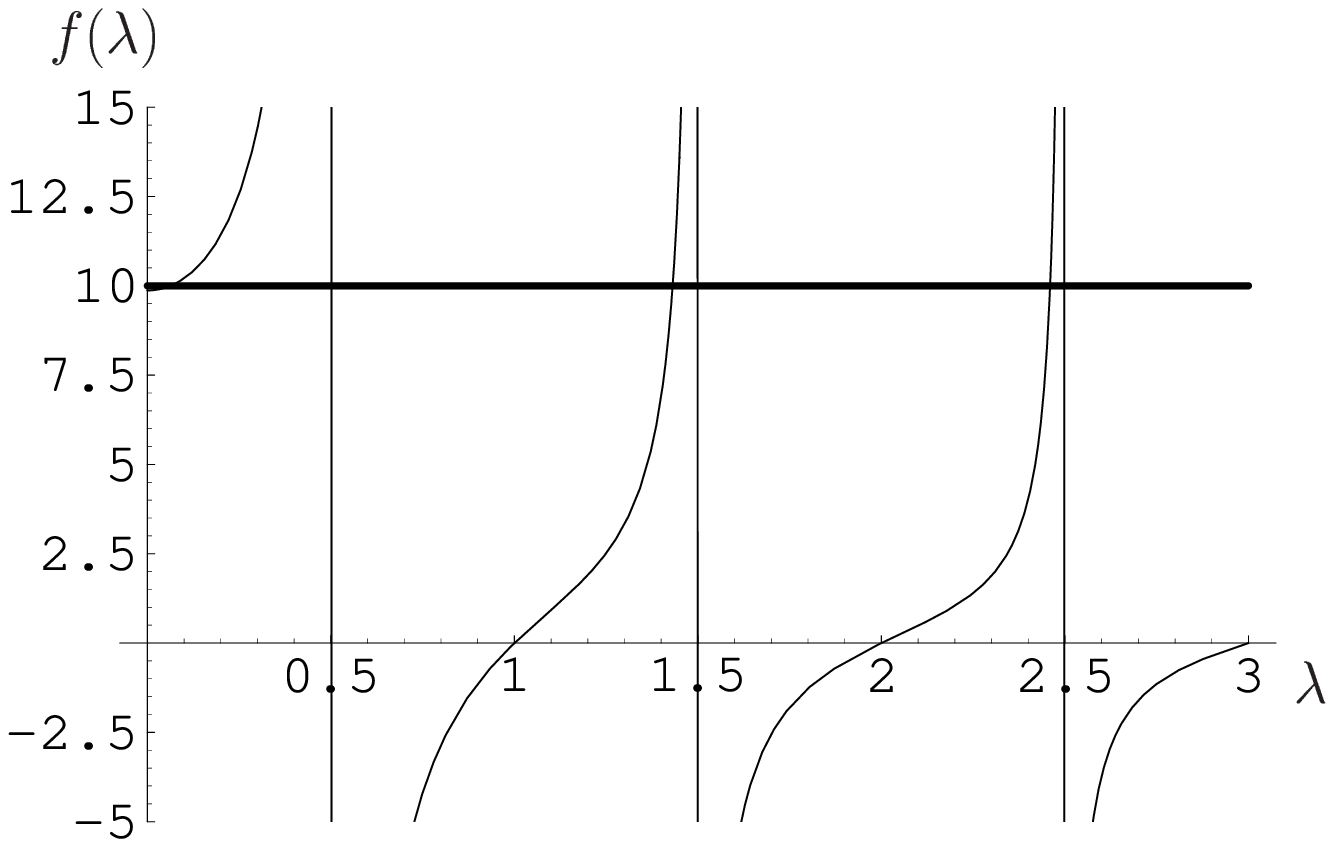}~
\includegraphics[width=6cm]{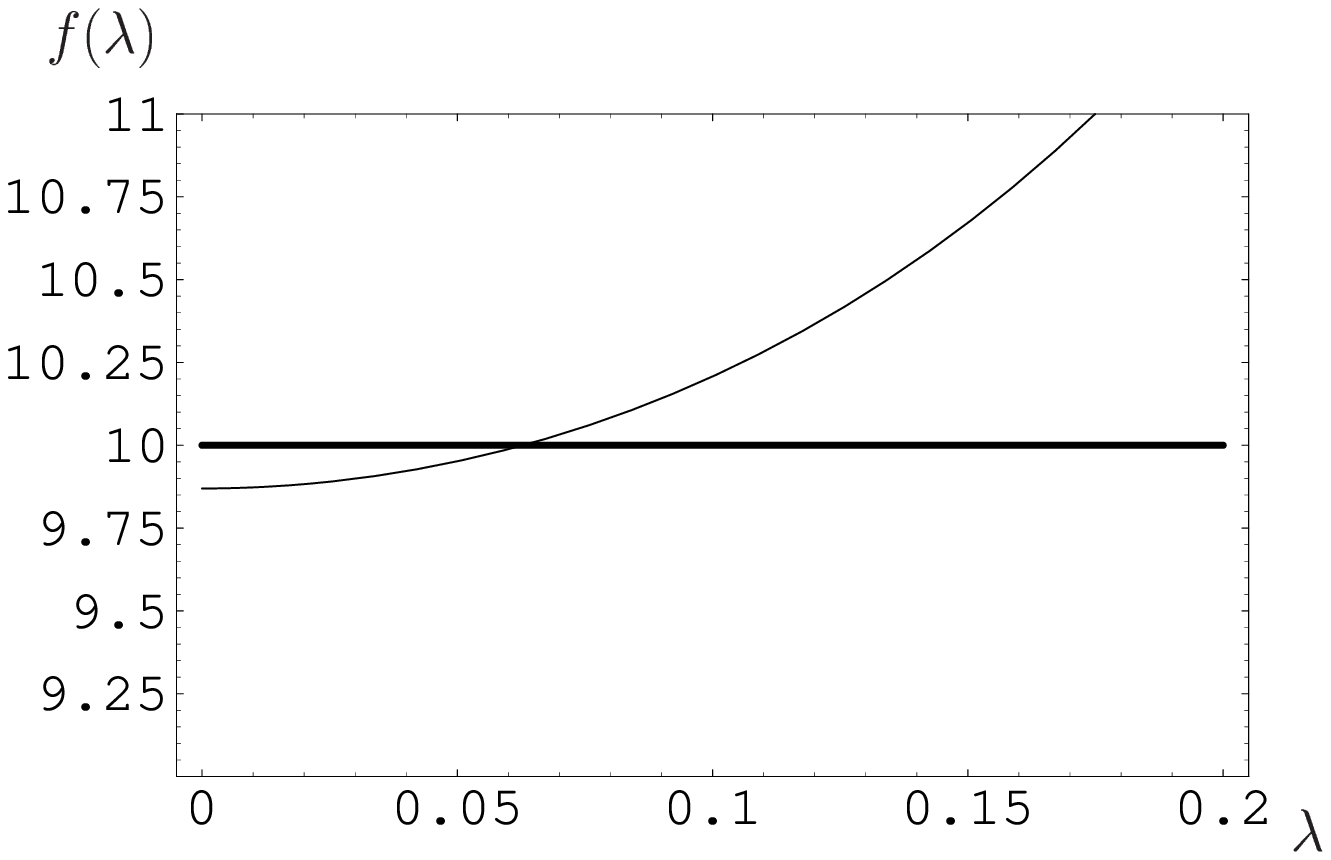} 
\caption{$f(\lambda)=\pi\lambda^{-1}\tan(\pi \lambda)$ (solid line)
and $f(x)=-1/P$ (bold line). For illustration, the graph is shown for $P=-0.1$.
In the right graph the neighborhood of the intersection for the lowest eigenvalue
is magnified. 
 \label{figb}}
\end{center}
\end{figure}

Now we examine $\lambda$ as a function of $P$ for $B=1/2$.
Using partial-fraction expansion of trigonometric functions
\bea
 \pi \tan \pi \lambda=
 - {2\lambda\over \lambda^2-
 {1\over 4} }
 -\sum_{n=1}^\infty {2\lambda\over \lambda^2-
 \left(n+{1\over 2}\right)^2 }
=  -\sum_{n=1}^\infty {2\lambda\over \lambda^2-
 \left(n-{1\over 2}\right)^2 } ,
\eea
from Eq.~(\ref{gomasseq}) we obtain 
\bea
   {\pi \tan\pi\lambda\over \lambda}=-{1\over P} .
  \label{weget}
\eea 
The function $\lambda^{-1}\tan \pi\lambda$ has any value from $-\infty$ to $\infty$
for $n-{1\over 2} <\lambda < n+{1\over 2}$ for each $n$.
Therefore, $\lambda$ in Eq.~(\ref{weget})
has multiple solutions correspondingly to Kaluza-Klein modes.
In Figure~\ref{figb}, behavior of each side of Eq.~(\ref{weget}) against
$\lambda$ is shown.
As seen in Figure~\ref{figb},
the mixing of $B$ and $|P|$ of order ${\cal O}(0.1)$ generates
$\lambda$ of order ${\cal O}(0.01)$. A smaller lowest-eigenvalue can be generated
by the mixing depending on the value of $P$.
For a very small $\lambda$, 
with the expansion of $\tan (\pi\lambda)=(\pi \lambda)+(\pi \lambda)^3/3+\cdots$,
Eq.~(\ref{weget}) is
\bea
   \pi^2 +{\pi^4 \lambda^2\over 3} +\cdots =-{1\over P} ,
\eea
up to ${\cal O}(\lambda^4)$.
To the order $\lambda^2$,
the solution of the lowest-eigenvalue is obtained as
\bea
   \lambda^2 =-{3\over \pi^4} 
       \left({1\over P}+\pi^2\right) .
       \label{lows}
\eea
Because the left-hand side is positive, there is a lower bound to $P$,
\bea
    P > -{1\over \pi^2} .
\eea
The existence of such a lower bound is seen in Figure~\ref{figa}.

In order to analyze the others of multiple solutions,
we take the $n$-th eigenvalue (except for the lowest (\ref{lows})) as $\lambda_n=n+{1\over 2} -\Delta_n$ 
to solve a small $\Delta_n$ perturbatively.
The perturbation is valid especially for a large $n$ because 
$\Delta_n/n$ is very small for a large $n$. 
With this normalization, Eq.~(\ref{weget}) is 
\bea
   \tan(\Delta_n\pi) = -{P \over n+\textrm{${1\over 2}$} -\Delta_n} .
    \label{befo}
\eea
By the expansion $\tan (\Delta_n\pi)=(\Delta_n\pi)+(\Delta_n\pi)^3/3+\cdots$,
Eq.~(\ref{befo}) is
\bea
   \Delta_n^2 -(n+\textrm{${1\over 2}$})\Delta_n -\textrm{${P\over \pi}$} =0 ,
\eea
up to ${\cal O}(\Delta_n^3)$.
We obtain the solution for $\Delta_n$,
\bea
   \Delta_n &\!\!\!=\!\!\!&
  {1\over 2}\left[(n+\textrm{${1\over 2}$})
    \pm \sqrt{(n+\textrm{${1\over 2}$})^2 + \textrm{${4P\over \pi}$}}
     \right]
\nonumber
\\
 &\!\!\!=\!\!\!&
   -{P\over (n+\textrm{${1\over 2}$})\pi}
   +{P^2\over (n+\textrm{${1\over 2}$})^3\pi^2} 
   +\cdots ,
   \label{sol}
\eea
where the ellipses denote higher order terms in the expansion 
with $P/(n+{1\over 2})^2$.
The minus sign in the first line of Eq.~(\ref{sol}) has been chosen 
so that the equation at the first-order of $\Delta_n$ is fulfilled.

Within the above approximation, we obtain the stop mass-eigenvalues squared
\bea
  m_{\tilde{t}}^2
  =-{3\over R^2 \pi^4}\left(
  {1\over P}+\pi^2\right) , ~~
  {1\over R^2}\left(n+\textrm{${1\over 2}$}
    +{P\over (n+\textrm{${1\over 2}$})\pi}\right)^2 ,  
    \label{stbp}
\eea
for $B=1/2$.
For a generic $n$,
the Kaluza-Klein mass is obtained as
\bea
 m_{\tilde{t}}^2
  = {1\over R^2}\left(n+  \Delta \right)^2 , ~~\Delta \approx 0.4 -0.5 .  
    \label{stbp2}
\eea
as found in Figure~\ref{figb}.

\subsubsection*{The Higgs boson mass corrections}

In Eq.~(\ref{mhlbk}) in Section~\ref{rad}, we wrote down the upper bound 
involving Kaluza-Klein modes.
The analysis  was based on
the simplification that
the $n$-th top and stop Kaluza-Klein masses are $n^2/R^2$ and
($m_{\tilde{t}}^2 +n^2/R^2$), respectively.
This structure of Kaluza-Klein masses are somehow different from
the result given in Eqs.~(\ref{stbp}) and (\ref{stbp2}). 
We here derive the Higgs boson mass corrections
for the lightest top and stop masses
\bea
   m_{t}^2 = |y_t|^2 v_2^2,\quad 
   m_{\tilde{t}}^2= -{3\over R^2 \pi^4}\left({1\over P}+\pi^2\right) ,
    \label{lit}
\eea
respectively and the $n$-th Kaluza-Klein top and stop masses
\bea
   m_{t,n}^2= {n^2\over R^2} , \quad
   m_{\tilde{t},n}^2 \approx
  {(n+\textrm{${1\over 2}$})^2 \over R^2} ,
     \label{torn}
\eea 
respectively.
For the lightest top and stop with the masses (\ref{lit}),
the correction to the Higgs boson mass is of the form (\ref{contz}),
\bea
  \Delta_t 
 ={3\sqrt{2}m_t^4 G_F\over 2\pi^2 \sin^2\beta}
 \ln\left({m_{\tilde{t}}^2\over m_t^2}\right) , 
  \label{zerod}
\eea
where $m_t$ and $m_{\tilde{t}}$ are given in Eq.~(\ref{lit}).   
For the Kaluza-Klein mode, we adopt a calculation technique 
employed in Ref.~\cite{Delgado:1998qr} in which
the potential 
\bea
  V={1\over 2} \textrm{Tr}
    \int {d^4p \over (2\pi)^4}
     \sum_{n=-\infty}^{\infty}
      \ln 
      \left({p^2 +M^2(\phi) +(n+q_b)^2/R^2
      \over p^2 +M^2(\phi) +(n+q_f)^2/R^2} \right) ,
\eea
is led to
\bea
  V&\!\!\!=\!\!\!&{1\over 128\pi^6 R^4}
    \textrm{Tr} \left[V(q_f, \phi) -V(q_b,\phi)\right] ,
\\
\!\!\!\! \!\!\!\!
 V(q ,\phi) &\!\!\!=\!\!\!&
 (2\pi R)^2 M^2(\phi) \textrm{Li}_3 
  (e^{2i\pi q} e^{-2\pi R\sqrt{M^2(\phi)}})
\nonumber
\\
   && +6\pi R\sqrt{M^2(\phi)}\textrm{Li}_4 
  (e^{2i\pi q}e^{-2\pi R\sqrt{M^2(\phi)}})
  +3 \textrm{Li}_5 (e^{2i\pi q}e^{-2\pi R\sqrt{M^2(\phi)}}) + \textrm{H.c.} .
\eea
where
$\textrm{Li}_n (z)=\sum_{k=1}^\infty  {z^k/k^n}$.
While the zero-mode potential 
(\ref{pou}) is
\bea
  U_t = 6
    \int {d^4p \over (2\pi)^4}
      \ln 
      \left({p^2 + m_{\tilde{t}}^2 +|y_t H_2^0|^2
      \over p^2 + |y_t H_2^0|^2} \right) ,
\eea
with the stop mass (\ref{lit}),
the potential contributed from Kaluza-Klein modes with the masses (\ref{torn})
can be written as
\bea
  U_{t\,\textrm{\scriptsize KK}}
  &\!\!\!=\!\!\!& U_{\textrm{\scriptsize sum}}
    -U_{\textrm{\scriptsize sub}} ,
\\
  U_{\textrm{\scriptsize sum}} &\!\!\!=\!\!\!& 6
    \int {d^4p \over (2\pi)^4}
     \sum_{n=-\infty}^{\infty}
      \ln 
      \left({p^2 +|y_t H_2^0|^2 + (n+\textrm{${1\over 2}$})^2/R^2
      \over p^2 + |y_t H_2^0|^2 +n^2/R^2} \right) ,
      \label{4.83}
\\
 U_{\textrm{\scriptsize sub}} &\!\!\!=\!\!\!& 6
    \int {d^4p \over (2\pi)^4}
      \ln 
      \left({p^2 +|y_t H_2^0|^2 +1/(4R^2)
      \over p^2 + |y_t H_2^0|^2} \right) .
\eea
Using the formula
\bea
  V(q_b=1/2,\phi) &\!\!\!=\!\!\!&
     -{(2\pi R)^4 M^4(\phi)\over 4} \ln 2 +\cdots ,
\\
 V(q_f=0, \phi) &\!\!\!=\!\!\!& 
  \left({3\over 16}-{1\over 4}\ln (2\pi R \sqrt{M^2(\phi)})\right)
     (2\pi R)^4 M^4(\phi) +\cdots ,
\eea
up to $M^2(\phi)$-linear term and
higher order terms ${\cal O}((RM(\phi))^{5/2})$,
we find the corrections for the Higgs boson mass,
\bea
   \Delta_{\textrm{\scriptsize sum}}
     =2v_2^2 U_{\textrm{\scriptsize sum}}'' (v_2^2)
     &\!\!\!=\!\!\!&-{3\over 2\pi^2} {m_t^4\over v_2^2} \ln (\pi R m_t) ,
\\
  \Delta_{\textrm{\scriptsize sub}}
   =2v_2^2 U_{\textrm{\scriptsize sub}}'' (v_2^2)
  &\!\!\!=\!\!\!&
         -{3\over 2\pi^2} {m_t^4\over v_2^2} \ln (2 R m_t) .
\eea
Thus the Kaluza-Klein mode corrections are
\bea
  \Delta_{t\, \textrm{\scriptsize KK}}
    =\Delta_{\textrm{\scriptsize sum}}
      -\Delta_{\textrm{\scriptsize sub}}
   ={3\sqrt{2} m_t^4 G_F\over \pi^2 \sin^2\beta} \ln {2\over \pi} .
   \label{kkdelt}
\eea
From this equation and the zero-mode contribution (\ref{zerod}),
we obtain the upper bound to the lightest Higgs boson mass,
\bea
   m_{h^0}^2 \leq 
    m_Z \cos^2 2\beta
     +{3\sqrt{2} m_t^4 G_F\over 2\pi^2}
       \ln {(2m_{\tilde{t}})^2\over (\pi m_t)^2}
 ,
       \label{final}
\eea
with $m_{\tilde{t}}^2$ given in Eq.~(\ref{lit}),
\bea
   m_{\tilde{t}}^2= -{3\over R^2 \pi^4}\left({1\over P}+\pi^2\right) .
   \label{4.91}
\eea
In Eq.~(\ref{final}), the Kaluza-Klein mode contribution is independent of $R$
and only $R$-dependence arises from the contribution of
the lightest top and stop.
The factor $\ln(m_{\tilde{t}}^2/m_t^2)$ must not be too large for some $R$.
For $B=1/2$,  a large $R^{-1}$ can be allowed if $P^{-1}\sim -\pi^2$.

Here we estimate the values of radiative corrections.
For $m_t =181~\textrm{GeV}$ and $m_{\tilde{t}}\sim 1000~\textrm{GeV}$
(composed of various values of $R$ and $P$ via Eq.~(\ref{4.91})),
the lightest mode contribution gives
\bea
   {3\sqrt{2} m_t^4 G_F\over 2\pi^2}
      \ln\left({m_{\tilde{t}}^2\over m_t^2}\right) \approx (96\textrm{GeV})^2 .
      \label{litc}
\eea
On the other hand, the Kaluza-Klein contribution makes
\bea
  {3\sqrt{2} m_t^4 G_F\over 2\pi^2}
       \ln  {4\over \pi^2}
  \approx -(50~\textrm{GeV})^2 .
\eea
Such a negative Kaluza-Klein contribution can be generated because 
in Eq.~(\ref{4.83}) the argument of the logarithm function
is smaller than unity for $n<0$.
These contributions 
$(96\textrm{GeV})^2-(50\textrm{GeV})^2=(82\textrm{GeV})^2$
can change the upper bound to the mass of the lightest Higgs boson
into an allowed region.

In the present model, gauge bosons propagate in bulk.
The gauge coupling is sensitive to energies over the inverse compactification scale~%
\cite{Dienes:1998vh}%
-\cite{Uekusa:2008iz}.
As a fundamental theory appears
before gauge couplings become very small or very large,
the number of Kaluza-Klein modes can be of at most ${\cal O}(100)$. 
Thus an ultraviolet momentum cutoff may be introduced in the theory.
Our result for the radiative correction to the Higgs boson mass is 
independent of such a cutoff.

\section{Conclusion \label{conclusion}}

We have studied corrections to the mass of the lightest Higgs boson
in a Kaluza-Klein effective theory.
For the simple spectrum, it has been explicitly shown how contributions of Kaluza-Klein modes to
the Higgs boson mass correction are summed.
There the summation over infinite numbers of Kaluza-Klein mode contributions
converges.
In other cases, a finite radiative correction has been obtained
using a calculation technique employed in the context with the summation over 
zero mode and Kaluza-Klein modes and a momentum integral.

While $N=1$ supersymmetry is oriented differently on each brane
depending on a Scherk-Schwarz twist,
an $F$-term on a brane has been introduced.
For these two sources, stop mass matrix is 
non-diagonal with respect to Kaluza-Klein modes and also
with respect to SU(2)${}_R$.
We have pointed out that a very small eigenvalue can be 
accommodated in the diagonalized basis.
Thus the mass splitting between zero-mode stop and top can 
be around $1~$TeV even for a large $R^{-1}$.
A large inverse compactification scale may be motivated by
four-dimensional MSSM gauge coupling unification.
It has also been shown that the eigenvalue equation becomes a simple form
with trigonometric functions for the Scherk-Schwarz parameter $\beta=\pi$.
In this case we have presented 
multiple solutions that correspond to Kaluza-Klein modes.
With these solutions, we have estimated 
radiative corrections to the lightest Higgs boson mass.

In the context of extra dimensions,
the mass splitting may be related to the inverse scale of compactification.
The stop mass can be 
\bea
   m_{\tilde{t}} \sim c \times {1\over R} ,
\eea
where $c$ is a numerical factor.
On the other hand, Kaluza-Klein masses are proportional to $R^{-1}$.
From these properties, the correction of Kaluza-Klein modes to the lightest Higgs boson mass
squared is independent of $R$
as in Eqs.~(\ref{corecn}) and (\ref{kkdelt}).
This Kaluza-Klein contribution can be around
over $3m_t^4 G_F/(\sqrt{2}\pi^2)= (52~\textrm{GeV})^2$. 
Together with zero-mode contribution,
it lifts the upper bound to the lightest Higgs boson 
in an allowed region,
although the value of zero mode contribution depends on $R$.

Finally, we mention stabilization of the scale $R$.
A simple setup to stabilize compactification radius is
to take a constant superpotential into account in a warped extra dimension%
~\cite{Maru:2006id, Maru:2006ji, Uekusa:2007sw}.
It may be interesting to 
examine the issue of the Higgs boson mass
for the mixing of two supersymmetry breaking sources in a warped 
model with radius stabilization.

\vspace{8ex}

\subsubsection*{Acknowledgments}

I am grateful to Masud Chaichian for a careful reading of the manuscript.

 


\vspace{1ex}



\end{document}